\documentclass[aps,pra,reprint,showpacs,nofootinbib,superscriptaddress]{revtex4-1}
\usepackage{graphicx, comment}
\usepackage{amsfonts}
\usepackage{amsmath,amssymb}
\usepackage{bm}
\usepackage{color}
\usepackage{epstopdf}
\usepackage{epsfig}
\usepackage{verbatim}
\usepackage{lineno}
\usepackage[thicklines]{cancel}
\usepackage{url}   
\usepackage{xcolor}
\usepackage{listings}
\usepackage{algorithm,algorithmic}
\usepackage{enumerate}
\usepackage{siunitx}
\usepackage{booktabs}
\usepackage{hyperref}
\hypersetup{
     colorlinks   = true,
     citecolor    = blue
}
\newcommand{\ket}[1]{\left| #1 \right>} 
\newcommand{\bra}[1]{\left< #1 \right|} 

\definecolor{red}{rgb}{1,0.,0}

\begin{document}

\title{Quantum Chemistry as a Benchmark for Near-Term Quantum Computers}

\thanks{This manuscript has been authored by UT-Battelle, LLC under Contract No. DE-AC05-00OR22725 with the U.S. Department of Energy. The United States Government retains and the publisher, by accepting the article for publication, acknowledges that the United States Government retains a non-exclusive, paid-up, irrevocable, world-wide license to publish or reproduce the published form of this manuscript, or allow others to do so, for United States Government purposes. The Department of Energy will provide public access to these results of federally sponsored research in accordance with the DOE Public Access Plan
(http://energy.gov/downloads/doe-public-access-plan).}

\author{Alexander J.\ McCaskey}
\email{mccaskeyaj@ornl.gov}
\affiliation{Quantum Computing Institute,\ Oak\ Ridge\ National\ Laboratory,\
  Oak\ Ridge,\ TN,\ 37831,\ USA}
\affiliation{Computer Science and Mathematics Division,\ Oak\ Ridge\ National\ Laboratory,\ Oak\ Ridge,\ TN,\ 37831,\ USA}
  
\author{Zachary P.\ Parks}
\affiliation{Quantum Computing Institute,\ Oak\ Ridge\ National\ Laboratory,\
  Oak\ Ridge,\ TN,\ 37831,\ USA}
\affiliation{Computer Science and Mathematics Division,\ Oak\ Ridge\ National\ Laboratory,\ Oak\ Ridge,\ TN,\ 37831,\ USA}
  
\author{Jacek Jakowski}
\email{jakowskij@ornl.gov}
\affiliation{Quantum Computing Institute,\ Oak\ Ridge\ National\ Laboratory,\
  Oak\ Ridge,\ TN,\ 37831,\ USA}
\affiliation{Computational Sciences and Engineering Division,\ Oak\ Ridge\ National\ Laboratory,\
Oak\ Ridge,\ TN,\ 37831,\ USA}
  
\author{Shirley V.\ Moore}
\affiliation{Quantum Computing Institute,\ Oak\ Ridge\ National\ Laboratory,\
  Oak\ Ridge,\ TN,\ 37831,\ USA}
\affiliation{Computer Science and Mathematics Division,\ Oak\ Ridge\ National\ Laboratory,\ Oak\ Ridge,\ TN,\ 37831,\ USA}

\author{T.\ Morris}
\affiliation{Quantum Computing Institute,\ Oak\ Ridge\ National\ Laboratory,\
  Oak\ Ridge,\ TN,\ 37831,\ USA}
\affiliation{Physics Division, \ Oak\ Ridge\ National\ Laboratory,\
  Oak\ Ridge,\ TN,\ 37831,\ USA}

\author{Travis S.\ Humble}
\affiliation{Quantum Computing Institute,\ Oak\ Ridge\ National\ Laboratory,\
  Oak\ Ridge,\ TN,\ 37831,\ USA}
\affiliation{Computational Sciences and Engineering Division,\ Oak\ Ridge\ National\ Laboratory,\
Oak\ Ridge,\ TN,\ 37831,\ USA}

\author{Raphael C.\ Pooser}
\email{pooserrc@ornl.gov}
\affiliation{Quantum Computing Institute,\ Oak\ Ridge\ National\ Laboratory,\
  Oak\ Ridge,\ TN,\ 37831,\ USA}
\affiliation{Computational Sciences and Engineering Division,\ Oak\ Ridge\ National\ Laboratory,\
Oak\ Ridge,\ TN,\ 37831,\ USA}
\date{\today}

\begin{abstract}
We present a quantum chemistry benchmark for noisy intermediate-scale quantum computers that leverages the variational quantum eigensolver, active space reduction, a reduced unitary coupled cluster ansatz, and reduced density purification as error mitigation. We demonstrate this benchmark on the 20 qubit IBM Tokyo and 16 qubit Rigetti Aspen processors via the simulation of alkali metal hydrides (NaH, KH, RbH),with accuracy of the computed ground state energy serving as the primary benchmark metric. We further parameterize this benchmark suite on the trial circuit type, the level of symmetry reduction, and error mitigation strategies. 
Our results demonstrate the characteristically high noise level present in near-term superconducting hardware, but provide a relevant baseline for future improvement of the underlying hardware, and a means for comparison across near-term hardware types. 
We also demonstrate how to reduce the noise in post processing with specific error mitigation techniques. Particularly, the adaptation of McWeeny purification 
of noisy density matrices dramatically improves accuracy of quantum computations,  which, along with adjustable active space, significantly extends the range of  accessible molecular systems.
We demonstrate that for specific benchmark settings, the accuracy metric can reach chemical accuracy when computing over the cloud on certain quantum computers.
\end{abstract}

\maketitle

\section{Introduction}
Noisy intermediate-scale quantum (NISQ) devices have been used recently to demonstrate a variety of different small-scale quantum computations~\cite{omalley2016, linke2017experimental, kandala2017, Dumitrescu2018, Klco2018}.
These demonstrations underscore the progress in developing programmable quantum processing units (QPUs) as well as advances in how to use these devices for scientifically meaningful computations. However, evaluating the impact of these demonstrations is complicated by the interplay of unique features available from different quantum computing technologies (superconducting electronics, trapped ions, etc.) and the multitude of programming choices that influence observed performance. Several properties quantify the intrinsic utility of a QPU including the capacity of the quantum register, the fidelity of the available instructions, the connectivity between register elements, etc.
While these hardware-specific parameters characterize progress in quantum hardware development, they do not directly benchmark the performance of a quantum computer for a given computational task. 
\par 
Benchmarks for application-specific metrics are needed to evaluate the efficiency and applicability of quantum computing for scientific applications.
In particular, comparing the benchmark performance of quantum computation to alternative computational approaches should rely on definitions that are meaningful across computational paradigms. Application-specific performance metrics have been used recently to evaluate the utility of near-term QPUs with respect to computational accuracy, including examples from machine learning \cite{Hamilton2018} and nuclear physics \cite{Dumitrescu2018}. However, the design and demonstration of an application-specific benchmark to evaluate QPU performance across scalable problem instances has been absent from the literature up to now.
\par 
We present a quantum chemistry simulation benchmark to evaluate the performance of quantum computing by defining a series of electronic structure calculation instances that can be realized on current hardware. We outline and investigate an array of parameter choices that influence quantum computational performance. Our approach uses multiple techniques including density matrix purification and active-space reduction via frozen-core approximation and truncation of the virtual space to accommodate hardware limitations such as a limited number of noisy qubits and a limited achievable circuit depth, while staying within the well-known hierarchy of quantum chemistry methods. For calculations on small molecular systems, such as alkali metal hydride molecules with multiple electrons presented here, the complexity of the quantum computation can be reduced to two valence electrons and the equivalent of a hydrogen molecule in minimal basis set. These approximations can be gradually lifted as the quality of quantum devices improves to provide a flexible benchmarking model.
\par
The quantum algorithmic primitive at the center of this benchmark is the state preparation circuit - the unitary coupled-cluster (UCC) ansatz~\cite{PhysRevA.95.020501} or hardware efficient (HWE) ansatz~\cite{kandala2017,Romero2018,Ryabinkin2018}, for example - used in the variational quantum eigensolver (VQE)~\cite{Aspuru-Guzik2005,omalley2016}, a noise-resilient algorithm which has been implemented on current noisy QPUs. This hybrid algorithm uses a classical search for the lowest eigenvalue of a given observable using the variational principle and the chosen ansatz, which is prepared and measured on the quantum computer. The utility of this algorithm has been demonstrated in a number of fields, including high-energy and nuclear physics \cite{klco_quantum-classical_2018, Dumitrescu2018} in addition to quantum chemistry \cite{Romero2018}. Its success is primarily due to the reliance on finite classical computing resources in tandem with invocations of the quantum computer for exponentially scaling aspects of the problem. We use multiple variations of the state-preparation primitives noted above in order to illustrates its effect on  performance and accuracy of quantum computations. The benchmark assesses accuracy of VQE for recovering the ground-state energy for a series of molecules when using different parameterized trial circuits and error mitigation strategies. The alkali metal hydrides currently included in the benchmark are NaH, KH, and RbH. We implemented calculations of the ground state energies of each of these molecules on the 20-qubit IBM Tokyo and 16-qubit Rigetti Aspen superconducting circuit devices. After a review of related work and quantum chemistry in sections II and III, we present a description of the benchmark in section IV, followed by the benchmark results in section V.

\section{Related Work}
The variational quantum eigensolver (VQE) has been used previously to calculate ground state energy \cite{peruzzo_variational_2014,omalley2016}, and it has become a leading example of quantum-classical computing. Several groups have demonstrated VQE applications using a variety of QPUs and trial circuits to target small molecular systems \cite{Romero2018,Kandala2018,Ryabinkin2018,omalley2016,Shen:2017, Dumitrescu2018,Klco2018}. VQE recovers the ground state energy of a Hamiltonian by searching for the quantum state that minimizes the observed energy expectation value. The complexity of this classical search through the parameter space depends strongly on the form of the underlying trial circuit. Optimization of the energy expectation value uses  preparation and measurement of the parameterized quantum circuit to drive the classical search method executed on a conventional computer. Existing demonstrations of VQE have revealed that despite the algorithm's noise resilience, error mitigation is required to overcome the extensive noise present in current platforms~\cite{Temme2017,Bonet:2018,Li2017}.

For chemistry applications, the main difference between HWE and UCC choices is that the latter provides a rigorous, electron number-conserving procedure for describing a trial wavefunction, but the corresponding circuit has a relatively large depth even for small systems. The HWE ansatz offers shorter circuit depth, but it supports a larger parameter space, and it prepares states with varying numbers of electrons that yield unphysical results.

In \cite{Romero2018}, the number of parameters, and hence the circuit depth, is reduced for the UCC ansatz by using pre-screening of cluster amplitudes. This pre-screening discards all the excitation operators for which the second order M\o ller-Plesset  perturbation amplitudes are below a chosen threshold. The number of qubits required for the VQE-UCC calculation, as well as the number of parameters required for preparation of the ansatz, are also reduced by using an active space approximation which divides the orbital space into a set of inactive and active orbitals with the occupation of the orbitals in the inactive space remaining unchanged.
In the present work 
we use a frozen core, second-quantized Hamiltonian approximation to further enable mapping of benchmark problems to noisy quantum hardware in a computational framework such as OpenFermion~\cite{open-fermion}.

\section{Chemistry on Quantum Computers}
Electronic structure calculations require a choice of basis set and simulation approach, which together have an impact on resource cost and accuracy of the simulation.
The hierarchy  of the  theoretical  methods and basis sets  has been established over the years. The most popular  methods  in order of increasing accuracy  (and computational cost)  are  Hartree-Fock (HF),  density  functional  theory (DFT),   perturbation theories (PT)  and configuration interaction (CI),   coupled cluster methods, full configuration interaction (FCI). 
Neither HF nor FCI is used  in any practical calculations since the former one is too inaccurate  while the latter one is too expensive.  Similarly, basis sets of increasing flexibility and size have been developed. The  two most  popular classes of basis functions are  Pople basis sets (for example 3-21G or 6-31G)  and Dunning basis sets (for example cc-pVDZ, cc-pVTZ). The smallest available basis set, also called the minimal basis set, is denoted as STO-3G (Slater Type Orbital contracted with 3 Gaussians). Although the STO-3G basis set is not used in practical classical calculations, it represents a good first model for development and benchmarking of new computational methods.

\begin{figure*}[ht]
\centering
\begin{tabular}{ll}
\includegraphics[width=0.6\textwidth]{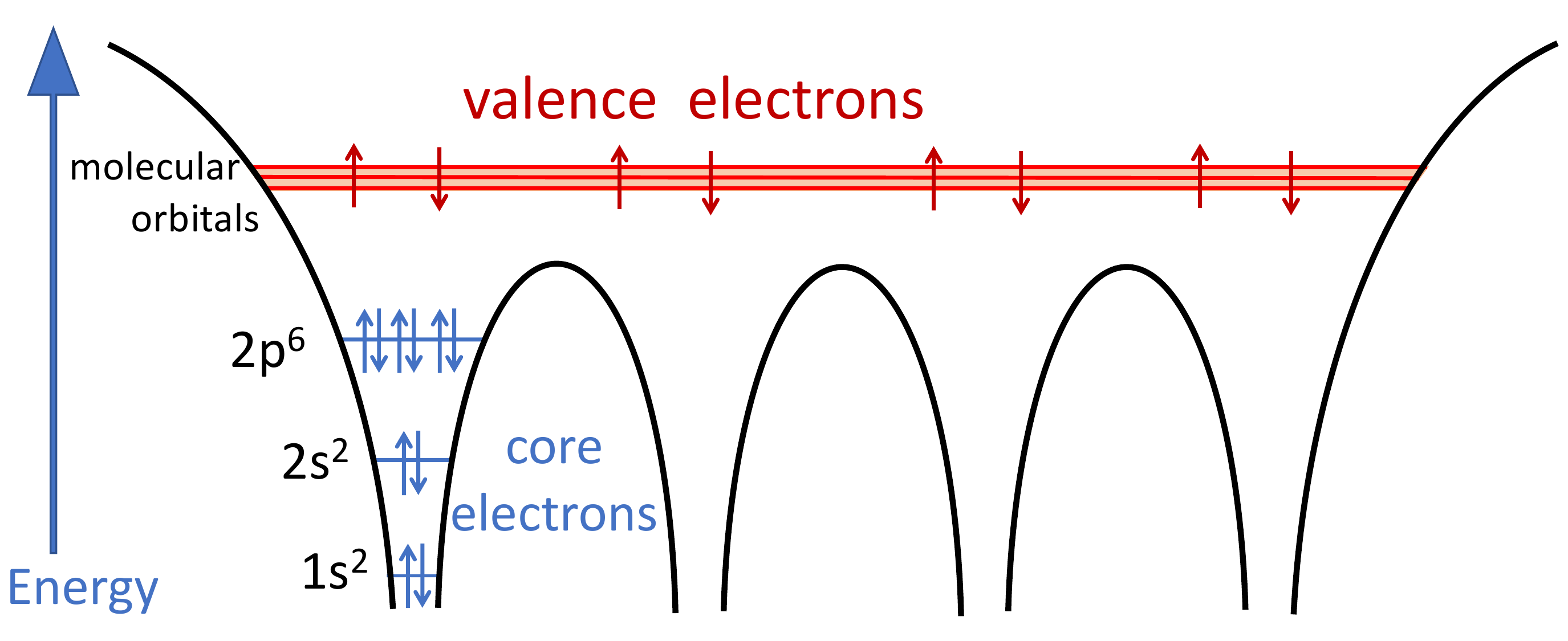} &
\includegraphics[width=0.2\textwidth]{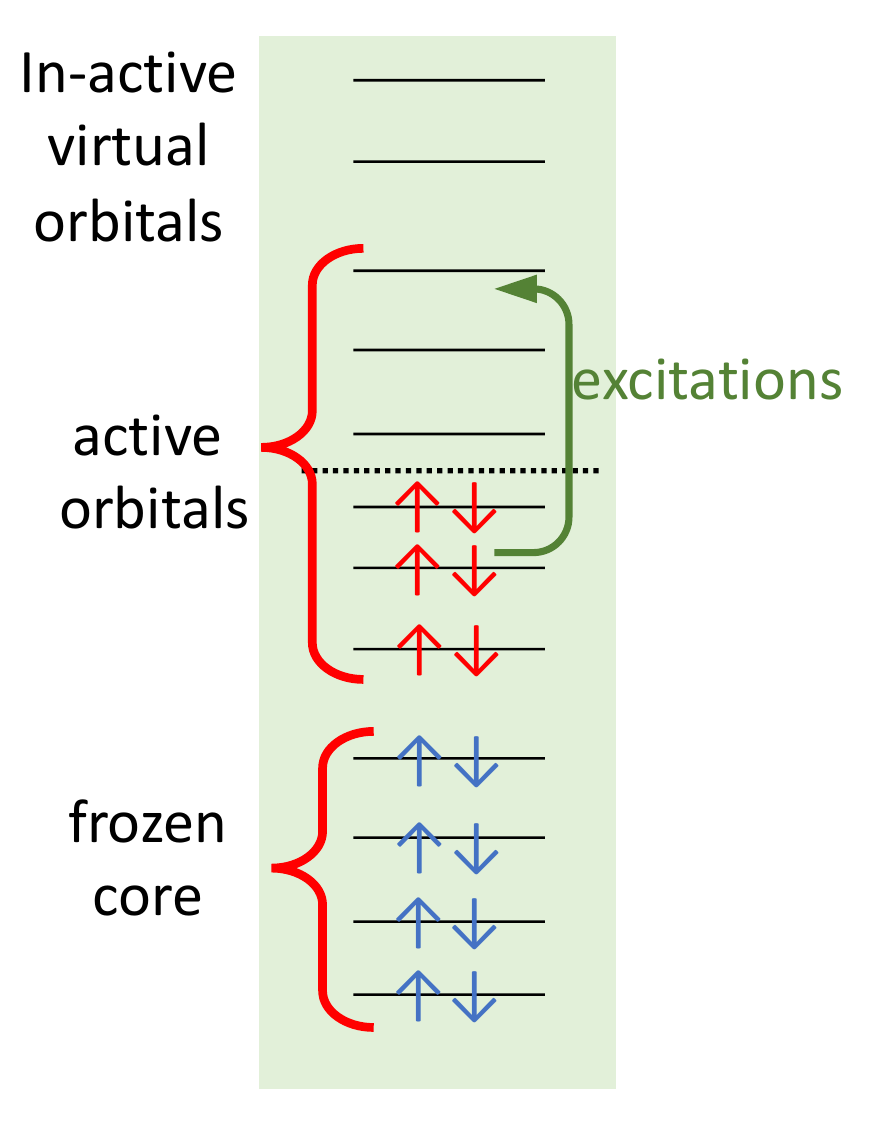} \\
(a) & (b) \\
\end{tabular}
\caption{ Frozen core approximation which partitions the orbital space into frozen core, active space, and inactive virtual space }
\label{fig:frozen-core}
\end{figure*}

Mapping the quantum chemistry problem to quantum computers involves  several steps: (1) molecule specification (structure, charge etc.);  (2) generation of the integrals and Hartree-Fock calculations to set an initial reference state; (3) transformation of the Hamiltonian from second quantization formalism to a qubit  representation with either the Jordan-Wigner or Bravyi-Kitaev transformation;  (4) generation of a quantum circuit that represents the trial wavefunction; and (5) mapping the quantum circuit to specific quantum computing hardware.

The electronic structure calculations in the benchmark seek the ground state energy configuration of the  Hamiltonian described as a sum of a nuclear repulsion  (0-electrons)  $H_0$  term, a one-electron term (kinetic energy of electrons plus interaction with core ions) $H_1$, and a two electron   Hamiltonian $H_2$
\begin{eqnarray}
\label{eq:H_tot}
\hat{H}&= & H_0 +H_1 +H_2 \\ \nonumber  
       &= & H_0 + \sum_{p,q} h^{p}_{q} \cdot  \hat p^\dagger \hat q   
         + \frac{1}{2} \sum_{p,q,r,s} g^{pq}_{sr}   \cdot  \hat{p}^\dagger \hat{q}^\dagger \hat{r} \hat{s},
\end{eqnarray}
where $p$, $q$, $r$, $s$ run over all molecular spin orbitals,  $\hat p^\dagger$,  $\hat q$ etc.\ are corresponding electron creation and annihilation operators, and  $h^{p}_{q}$  are matrix elements of the core Hamiltonian (kinetic energy of electrons plus interaction with core ions).  The $g^{pq}_{sr}$ are two-electron repulsion integrals $g^{pq}_{sr} = \langle p, q| s,r  \rangle$. The above  expression for the Hamiltonian  in the chemistry notation can be equivalently  written in the so called physicists notation:
\begin{eqnarray}
 \hat{H}      &= & H_0 + \sum_{p,q} h^{p}_{q} \cdot  \hat p^\dagger \hat q   
         + \frac{1}{4} \sum_{p,q,r,s} \bar g^{pq}_{sr}   \cdot  \hat{p}^\dagger \hat{q}^\dagger \hat{r} \hat{s},
\end{eqnarray}
with $\bar g^{pq}_{sr}$ being an anti-symmetrized repulsion  integral
\begin{equation}
\bar g^{pq}_{sr}=  g^{pq}_{sr}- g^{pq}_{r,s}=  \langle p, q| s,r  \rangle - \langle p, q| r,s  \rangle.   
\end{equation}

To reduce depth and enable execution on noisy near-term quantum computers, we neglect the contribution to electronic correlation from the lowest energy core electrons.  We treat only the interaction of correlated electrons with core electrons in an average mean-field fashion. Thus we generate the following reduced effective Hamiltonian:
\begin{equation}
H =  H'_0   +  H'_1  +H'_2, 
\end{equation}
with matrix elements $\mu,\nu$ given by
\begin{equation}
H'_{0} =  E_{nucl}  +\sum_a \big( {h}^a_a + \tfrac{1}{2} \sum_{b} \bar{g}^{ab}_{ab}\big),  
\end{equation}
where $a$ and $b$  run over frozen-core spin-orbitals.
Similarly,  the 1-body  term is given by
\begin{equation}
H'_{1} = \sum_{p,q}  \hat p^\dagger  \hat q   \cdot
\Big( {h}^{p}_q + \tfrac{1}{2}
\sum_a   \bar{g}^{a p }_{a q} \Big)  
\end{equation}
and the 2-body part is 
\begin{equation}
H'_2 = \tfrac{1}{4} \sum_{p,q,r,s}   \bar{g}^{pq}_{sr}  \cdot  \hat p^\dagger \hat q^\dagger \hat r \hat s.   
\end{equation}
Here we note that in the last  three expressions,  the indices $p$, $q$, $r$, $s$ run over active spin orbitals instead of all spin orbitals.  Indices $a$ and $b$ run over frozen core orbitals which are not active.  Finally, one can exclude   the specific   virtual orbitals from the active space.    Figure \ref{fig:frozen-core} illustrates partitioning  of orbital space  onto frozen core, active space  and inactive virtual space.

In order to calculate the ground state energy of Eq.~\ref{eq:H_tot} using VQE, one begins by mapping the fermionic representation to an equivalent spin-based Hamiltonian 
\begin{equation}
\begin{split}
    \label{eq:H_spin}
    \hat{H} = \sum_{i,\alpha} h^i_\alpha \sigma^i_\alpha + \sum_{i,j,\alpha,\beta} h^{ij}_{\alpha \beta} \sigma^i_\alpha \sigma^j_\beta \\
    + \sum_{i,j,k,\alpha,\beta,\gamma} h^{ijk}_{\alpha \beta \gamma} \sigma^i_\alpha \sigma^j_\beta \sigma^k_\gamma + \ldots
\end{split}
\end{equation}
via well-known transformations \cite{bk:2002,bk_jw:2012}. Here, Pauli tensor product $\sigma^i_\alpha \in \{\sigma_x,\sigma_y,\sigma_z,I\}$ 
acts on the $i^{th}$ qubit. Next, the qubit register is initialized to an unentangled reference state $\ket{\psi_0}$, followed by the application of a parameterized unitary $U(\bm{\theta})$ producing the state $\ket{\psi( \bm{\theta})} = U( \bm{\theta} )\ket{\psi_0}$. Measurements dictated by the Pauli tensor product terms in Eq. \ref{eq:H_spin} are then performed on this state to compute the individual Pauli term expectation values, which are then contracted with term coefficients to produce the variational energy ${\bar{E} (\bm{\theta})}  = \bra{\psi(\bm{\theta})}\hat{H} \ket{\psi(\bm{\theta})}$. The parameters $\bm{\theta}$ are then updated as part of a user-specified classical optimization routine and the entire process is repeated until convergence to the minimum of $E(\bm{\theta})$.

\section{Benchmarking Approach}
The VQE algorithm introduces a number of critical input generation decisions that are of interest to benchmarking efforts for quantum chemistry on NISQ hardware. Aspects of this algorithm that one may vary are the overall complexity of the spin Hamiltonian (i.e., consider symmetry-reduced versions of Eq. \ref{eq:H_spin}), the parameterized trial circuit $U(\bm{\theta})$, and the error mitigation strategy. These choices  dictate the complexity of the electronic structure problem that can be solved, and the accuracy, which is used as a benchmark metric. 

\begin{figure}
\centering
\includegraphics[width=\columnwidth]{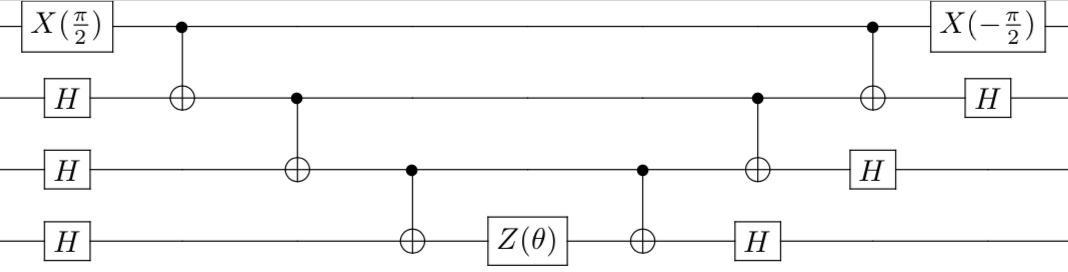}
\caption{The single parameter UCC (denoted ucc-1) circuit ansatz for 4 qubits and 2 electrons. It serves as a primitive circuit type that will be found as a sub-circuit in any UCC-type ansatz simulation, and therefore serves as an excellent benchmark for quantum chemistry on near-term quantum computers. A parameter $\theta$ controls a double excitation amplitude resulting in UCCD ansatz.}
\label{fig:uccsd}
\end{figure}

Our approach first seeks to reduce the complexity of the Hamiltonian in an effort to minimize the width of our quantum circuits. In addition to the frozen core approximation, we also reduce the number of qubits necessary for simulation via the application of discrete $\mathbb{Z}_2$ symmetries~\cite{tapering}. Next, we choose a particular trial unitary class. Finally, we consider our benchmarking results as a function of various error mitigation post-processing techniques. 

\subsection{Trial Circuits}
The choice of ansatz dictates the depth of the program executed and therefore the level of noise present in the computation. The number of parameters in the circuit determines the difficulty of the classical optimization step and the number of individual QPU calls required. 

In the UCC method the wavefunction is represented using the exponentiated operator~\cite{omalley2016,Romero2018,Taube-2006}
\begin{equation}
\ket{\psi(\bm{\theta})} = e^{T-T^\dagger}\ket{\psi_0}
\label{eq:ucc}
\end{equation}
where $\ket{\psi_0}$ is a reference state; for example, the Hartree-Fock solution.  The symbol  $T$ ($T^\dagger$) is a particle excitation (de-excitation) operator, given by
\begin{eqnarray}
T   &=&  \sum_{k=1}^{M}{T_k} (\bm{\theta}) \\
T_1(\bm{\theta}) &=& \sum_{\substack{i\in \text{occ} \\ a \in \text{virt}}} \theta_{a}^{i} a_{a}^\dagger a_{i} \\
T_2(\bm{\theta}) &=& \frac{1}{4}\sum_{\substack{i,j\in \text{occ} \\ {a,b} \in \text{virt}}} \theta_{a,b}^{i,j}  a_a^\dagger  a_b^\dagger  a_i a_j \\
&\ldots & \nonumber
\label{eq:ucc-amplitudes}
\end{eqnarray}
where $M$ is the number  of electrons  and the $\theta$ parameters map directly to the parameters 
found in the circuit decomposition of this unitary operator. For UCC single doubles (UCCSD) the excitation operator is 
truncated to single and double excitations only, $T=T_1(\bm{\theta}) + T_2(\bm{\theta})$. 
In general, the UCCSD ansatz is sufficient to map the exact FCI solution for 2-electron systems, such  as alkali hydrides within frozen core approximation. 
Note that for N spin-orbitals and M electrons, the number of parameters for the UCCSD circuit scales as $O(N^2M^2)$~\cite{Romero2018},
and therefore the depth for even modest sized problems grows well past what is implementable on currently available quantum computers. For a 4-qubit, 2-electron problem, the depth of this circuit is on the order of 100, and the number of instructions is about 150. Thus, symmetry reduction is required to implement the ansatz on current hardware.  

\begin{figure} 
\centering
\includegraphics[width=\columnwidth]{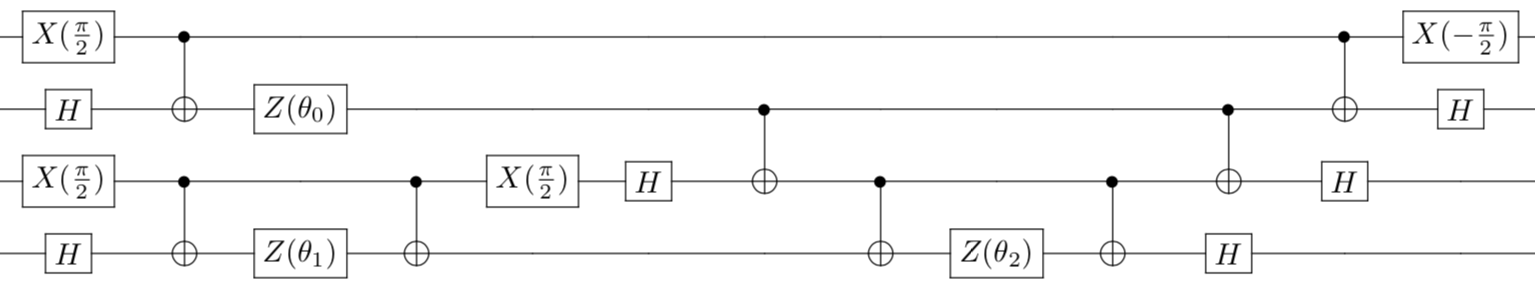}
\caption{The three parameter UCC (denoted ucc-3) circuit ansatz for 4 qubits and 2 electrons. This ansatz adds a parameter for each of two additional 2-qubit subspaces of the ucc-1 circuit ansatz so as to cover all active orbitals resulting in UCCSD circuit. The parameter $\theta_1$  ($\theta_2$) controls single excitation amplitude within  $\alpha$ spin ($\beta$ spin) block.}
\label{fig:uccsdex}
\end{figure}

One such reduction uses the fact that the spin terms that make up the exponential argument of Eq.~\ref{eq:ucc} for a hydrogenic problem (4 orbitals, 2 fermions) consist of 8 four-site terms that all act identically on the Hartree-Fock state. Therefore we can approximate the state with only one term. Here, we use $Y_0 X_1 X_2 X_3$ so that our parameterized unitary becomes
\begin{equation}
\label{eq:ruccsd}
    \hat{U}(\theta) = e^{i\theta Y_0 X_1 X_2 X_3}.
\end{equation}
The circuit for this unitary is shown in Figure \ref{fig:uccsd}. This circuit structure is a computational primitive that recurs in larger systems. This CX ladder structure is therefore a prime example of an early benchmark for quantum chemistry on quantum computers -- if the quantum device cannot adequately execute this circuit, it will perform poorly on all larger fermionic systems. On the other hand, one can map larger molecules onto this hydrogenic system via a frozen-core approximation, and apply this circuit primitive to produce energy metrics for a cross-platform comparison. 

We note that this ansatz cannot produce FCI energy for 4 qubit, 2 electron problems, but it provides a UCCD solution. To correct for this, we append circuits that address additional two-qubit subspaces in order to expand the total accessible Hilbert space represented by the quantum register (see Fig.~\ref{fig:uccsdex}). This extended UCC ansatz introduces an additional parameter that controls a single excitation for each two-qubit spin subspace, allowing for the electronic excitations to spread to fully cover all orbitals considered active in the Hamiltonian.

For completeness, we also consider an implementation of the hardware-efficient ansatz in \cite{kandala2017}. The circuit is composed of alternating layers of single-qubit rotations and two-qubit entangling operations (based on the connectivity structure of the hardware). The number of parameters in this circuit grows as $N(3d+2)$, where N is the number of qubits and d is the number of rotation+entangling layers. This ansatz is more amenable to near-term execution than a naive implementation of the UCC ansatz due to the controlled growth in its depth. However, the increase in the number of variational parameters leads to an increase in costly QPU executions (which are remote network calls for most available NISQ devices). Furthermore, this ansatz has been shown to introduce barren plateaus - regions where the probability that the gradient in a given direction is non-zero becomes exponentially small as the number of qubits increases \cite{mcclean_barren_2018}. 

\subsection{Error Mitigation Strategies}
Executing quantum circuits on current NISQ hardware will produce results that do not compare well with classical theoretical values. This is due to a number of factors, including intrinsic systematic noise present during execution and qubit measurement readout errors. Therefore, in order to produce valid results, some form of error mitigation must be employed, which introduces another variant into any quantum computing benchmarking approach. We examine three mitigation strategies: readout error mitigation, entangling gate error rate extrapolation, and reduced density matrix purification. 

We used a local, spatially uncorrelated readout error model that requires information about the probability of an unexpected bit flip during qubit measurement. For each qubit, we compute the probability that a $\ket{1}$ was observed when a $\ket{0}$ was expected, and the probability a $\ket{0}$ was observed when a $\ket{1}$ was expected, $p_i(1|0)$, $p_i(0|1)$, respectively. Then, our readout-error corrected expectation values are computed from the experimentally observed bit string counts as
\begin{equation}
    \left\langle Z \ldots Z \right\rangle = \sum_{x\in \text{counts}} p(x) \prod_{i\in \text{sites}(Z\ldots Z)}\left[ \frac{(-1)^{x_i} - p_i^-}{1-p_i^+}\right],
\end{equation}
where $p(x)$ is the probability of seeing bit string $x$ and $p_i^\pm = p_i(0|1) \pm p_i(1|0)$. $i \in \text{sites}(Z \ldots Z)$ represents the qubit indices represented in the given measured Pauli term. 

To mitigate against systematic two-qubit entangling gate noise, we implemented the zero-noise extrapolation technique put forth in \cite{Li2017}. We assume a generic two-qubit white noise error channel $\mathcal{N} = (1-\epsilon)\rho + \epsilon I/4$. We emulate increasing $\epsilon$ by introducing pairs of CNOT gates, serving as a noisy identity. We introduce $r$ pairs of these entangling operations, compute the energy produced by the circuit each time, and extrapolate back to $r=0$ to produce the noiseless energy.
  
Finally,  we adapted a McWeeny purification  scheme of non-idempotent density  matrices.
We implemented a mixed-state purification approach that depends on the computation of the two-body reduced density matrix (RDM)~\cite{Titus-in-prep}. Typically, VQE seeks the expectation value of an operator that can be expressed as a sum of individual, weighted Pauli tensor products. If we instead stay in the fermionic picture, we seek the minimum of the following energy expression
\begin{equation}
\begin{split}
    E(\bm{\theta}) = \bra{\psi(\bm{\theta})} H \ket{\psi(\bm{\theta})} = 
    \sum_{p,q} h_{pq} \bra{\psi(\bm{\theta})} a^\dagger_p a_q \ket{\psi(\bm{\theta})}  + \\
    \frac{1}{2} \sum_{p,q,r,s} h_{pqrs} \bra{\psi(\bm{\theta})} a^\dagger_p a^\dagger_q a_s a_r \ket{\psi(\bm{\theta})}, 
\end{split}
\end{equation}
where $\bra{\psi(\bm{\theta})} a^\dagger_p a^\dagger_q a_s a_r \ket{\psi(\bm{\theta})}$ is the two-body RDM (2-RDM), $\rho_{pqrs}$. From the study of \emph{n}-representability theory, an appropriate trace of the 2-RDM yields the one-body RDM. Therefore the 2-RDM is sufficient for computing the overall energy of the system~\cite{marginal}. We can compute these 2-RDM elements by constructing all physically-relevant $a^\dagger_p a^\dagger_q a_s a_r$ operators, mapping to the spin representation, executing the chosen parameterized ansatz, and performing measurements dictated by these transformed spin terms. This computation, when done in the presence of systematic noise, produces a statisical mixture of many pure states. Recall that the variational principle ensures that $ E(\bm{\theta}) \leq E_g$ for some pure state with corresponding eigenvalue $E_g$. We therefore employ a strategy for purifying these RDM elements, following the well-known McWeeny purification formula \cite{mcweeny}. 
To improve our estimate of $\rho_{pqrs}$ and therefore mitigate against inherent error on the hardware, we purify the 2-RDM via the iterative approach  
$\rho_{pqrs} \leftarrow  3(\rho_{pqrs})^2 - 2(\rho_{pqrs})^3$ until $\text{Tr}(\rho_{pqrs}^2 - \rho_{pqrs}) < \epsilon$, for $\epsilon \ll  1$. 
Note that here tensor multiplication is defined as the trace $C_{pquv} = A_{pqrs} B_{rsuv}$ (Einstein summation implied).

\subsection{Software Implementation}
Enabling a high-level, application-based benchmarking capability for near-term quantum computation requires an extensible and modular approach that abstracts away the underlying hardware and the benchmark application domain. The goal is to provide an executable that can be quickly and easily installed and an input deck that is expressive and enables one to tailor all available benchmark parameters. 

Our approach provides a hardware-independent benchmark that can be extended to new domains for application-centric and algorithmic primitive benchmarking. 
We have extended the Eclipse XACC quantum-classical programming framework \cite{xacc} and provided a Pythonic approach for designing and executing benchmarks across the major available QPUs. 
Our approach extends the Python API provided by XACC with support for runtime-contributed service interfaces for the various aspects of application-centric and primitive benchmarking. The benchmark takes an INI file as input.
This file describes the benchmark to be run, including the algorithm to use (VQE), input pertinent to the algorithm (trial circuit, optimizer, etc.), molecular integral generation routine, error mitigation strategies to implement (XACC automates the error mitigation strategies described in the previous section), and the target QPU to benchmark. A typical input file is shown in Listing~\ref{ini}. 
Providing a cross-platform benchmark capability is therefore a matter of distributing these input files for execution on currently available QPUs. Further details on the software framework are given elsewhere~\cite{xacc-bench-software}. 
\begin{lstlisting}[language=bash,caption={A typical benchmark input file  for NaH molecule with STO-3G basis set. 
Notice that  indexes 0 to 9 in the spin-orbitals lists  correspond  spin-up  orbitals whereas 10 to 19 correspond to  spind-down orbitals.}\label{ini}]
[XACC]
accelerator: ibm:ibmq_20_tokyo
algorithm: vqe

[Error Mitigation]
readout-error: True
richardson-extrapolation: True

[VQE]
optimizer: cobyla

[Ansatz]
name: ucc-3

[Molecule]
basis: sto-3g
geometry: 'Na  0.0 0.0 0.0
           H   0.0 0.0 1.914388'
frozen-spin-orbitals: [0,1,2,3,4, 
                       10,11,12,13,14]
active-spin-orbitals: [5,9,15,19]
\end{lstlisting} 
\begin{figure}
\begin{center}
\includegraphics[width=0.5\textwidth]{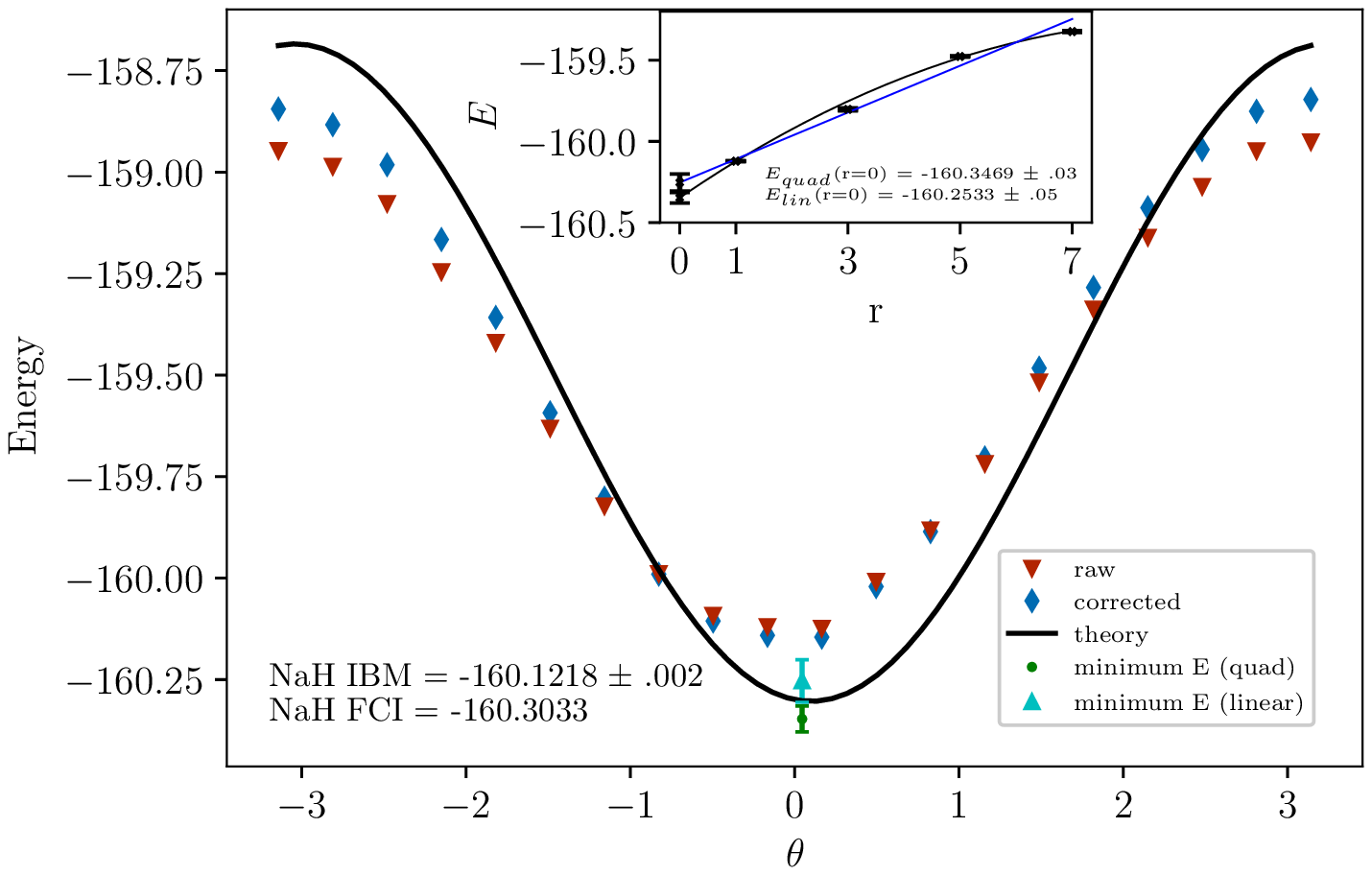} 
\includegraphics[width=0.5\textwidth]{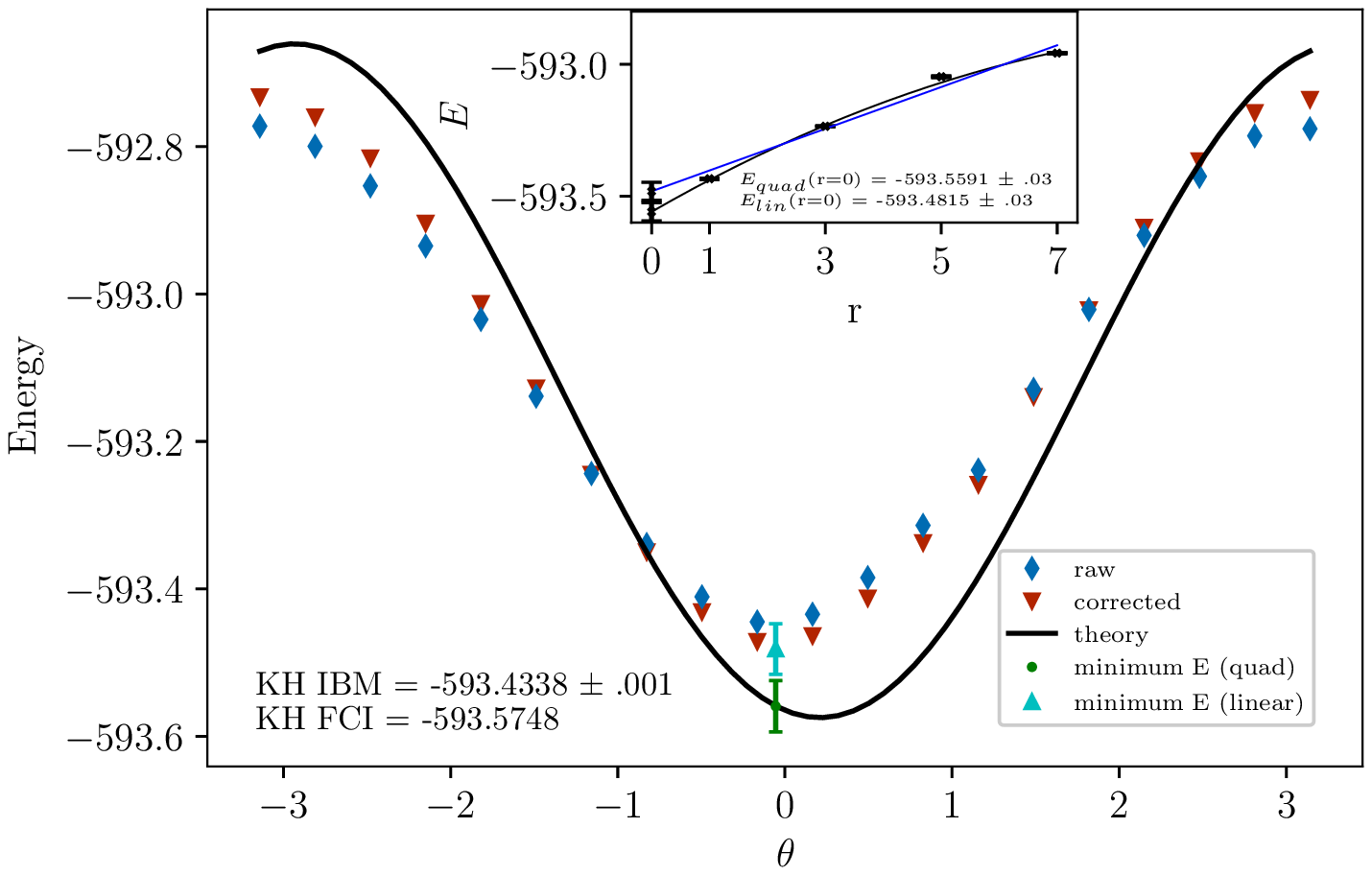} 
\includegraphics[width=0.5\textwidth]{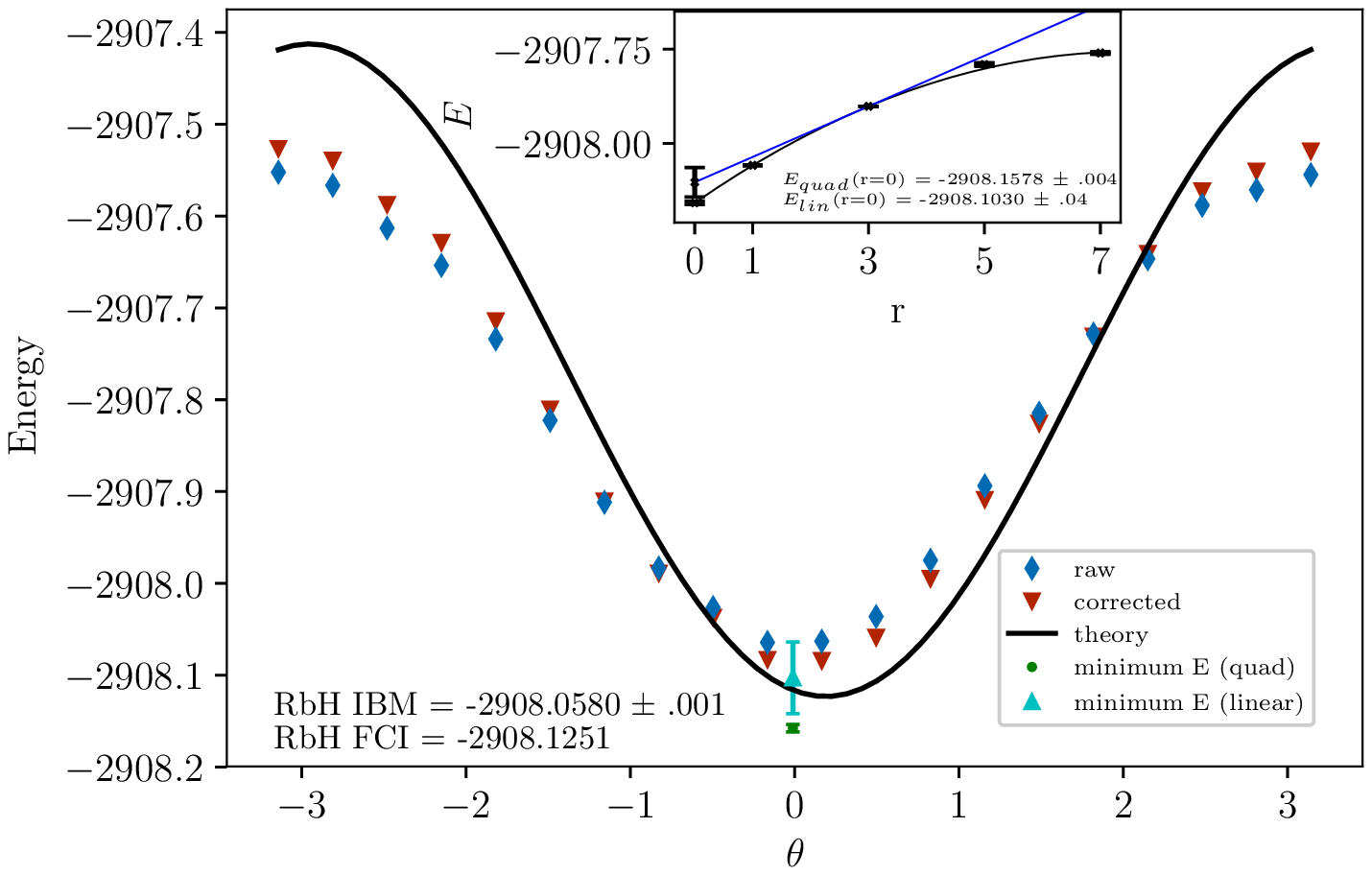} 
\end{center}
\caption{The dependence of the energy as a function of the variational parameter $\theta$ for the one parameter UCC ansatz for NaH (top), KH (middle), and RbH (bottom) on the 20-qubit IBM Tokyo QPU. The solid line corresponds to the theoretically exact $E(\theta)$ and the $\blacksquare$ ($\blacktriangledown$) markers correspond to raw (readout-error corrected) energies. The error bars for these points are smaller than the points and are on the order of $10^{-3}$. Linear and quadratic Richardson extrapolation results are displayed in the inset for NaH (top), KH (middle), and RbH (bottom) at optimal $\theta=0.04567616, -0.05485024, -0.01050553$, respectively. The teal and green data markers correspond to the minimum energies obtained from the linear and quadratic Richardson extrapolation, respectively.}
\label{fig:plots}
\end{figure}

\section{Results}
Our benchmark includes calculations for three different alkali metal hydrides -- NaH, KH, and RbH. For each of these molecules, we restrict the active space by freezing inner-core electrons, leaving four active orbitals with two fermionic degrees of freedom. We executed these benchmarks on the 20-qubit IBM Tokyo and the 16-qubit Rigetti Aspen QPUs. All fermionic Hamiltonians were mapped to spin Hamiltonians via the Jordan-Wigner transformation.

\begin{table}[b]
\caption{Computed ground state energies on IBM Tokyo}
\setlength\tabcolsep{0pt} 
\footnotesize\centering
This table details the computed energies (E) on the 20-qubit IBM Tokyo and the 16-qubit Rigetti Aspen QPUs for various alkali metal hydrides, ans{\"a}tze, and error mitigation strategies leveraged (EM - none, ro$+$re for readout-error and Richardson extrapolation, or rdm for RDM purification). The FCI energies for 4-qubit frozen-core NaH, RbH, and KH are $-160.3034597$, $-2908.125112$, and $-593.5747682$, respectively.
\smallskip 
\begin{tabular*}{\columnwidth}{@{\extracolsep{\fill}}rccccr}
\hline
\toprule
  Molecule & Ansatz & $Z_2$ Tapering & QPU & EM & E \\ 
\midrule
\hline
  NaH & ucc-1 & no & Tokyo & none  &  -160.122 $\pm$ 0.002 \\ 
  NaH & ucc-1 & no & Tokyo & ro+re &  -160.347 $\pm$ 0.032 \\ 
  NaH & ucc-1 & no & Aspen & none  &  -159.980 $\pm$ 0.004 \\
  NaH & ucc-1 & no & Aspen & rdm   &  -160.303 $\pm$ 0.004 \\
  NaH & ucc-1 & no & Tokyo & none  &  -159.973 $\pm$ 0.004 \\
  NaH & ucc-1 & no & Tokyo & rdm   &  -160.279 $\pm$ 0.004 \\
  NaH & ucc-3 & no & Aspen & none  &  -159.917 $\pm$ 0.013 \\
  NaH & ucc-3 & no & Aspen & rdm   &  -160.301 $\pm$ 0.013 \\
  NaH & ucc-3 & no & Tokyo & none  &  -160.049 $\pm$ 0.005 \\
  NaH & ucc-3 & no & Tokyo & rdm   &  -160.297 $\pm$ 0.005 \\
  NaH & hwe  & yes & Tokyo & none  &  -160.263 $\pm$ 0.009 \\ 
  NaH & hwe  & yes & Tokyo & ro+re &  -160.287 $\pm$ 0.002 \\ 
  NaH & hwe  & no & Tokyo  & none  &  -160.037 $\pm$ 0.005 \\ 
  NaH & hwe  & no  & Tokyo & ro+re &  -160.085 $\pm$ 0.007 \\ 
  KH  & ucc-1 & no & Tokyo & no    &  -593.434 $\pm$ 0.002 \\ 
  KH  & ucc-1 & no & Tokyo & ro+re &  -593.559 $\pm$ 0.035 \\ 
  RbH & ucc-1 & no & Tokyo & none  & -2908.058 $\pm$ 0.001 \\ 
  RbH & ucc-1 & no & Tokyo & ro+re & -2908.158 $\pm$ 0.004 \\ 
  \hline
\end{tabular*}
\label{tab:data}
\end{table}

\begin{figure}[ht!]
\begin{center}
\includegraphics[width=0.5\textwidth]{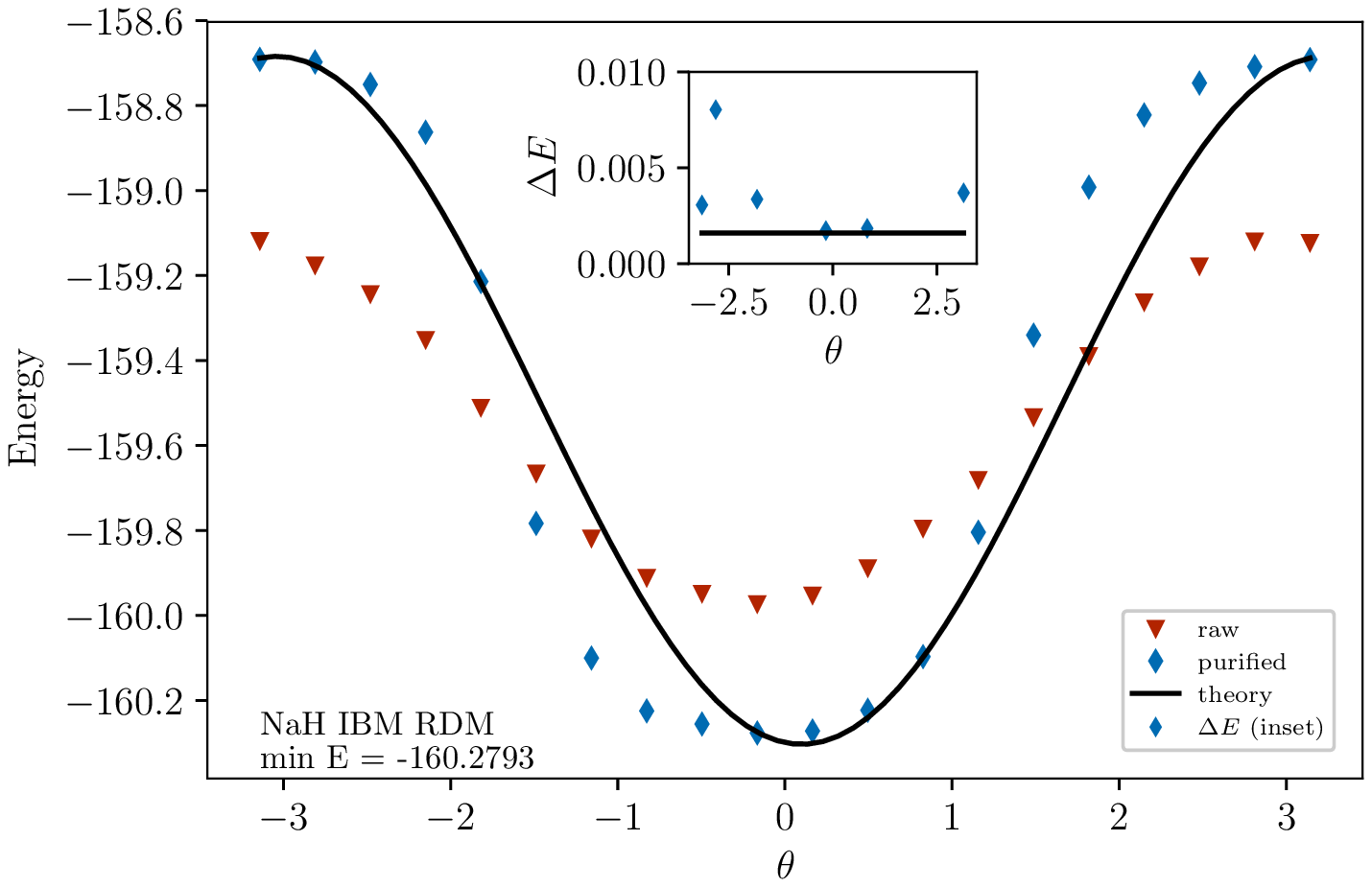} 
\includegraphics[width=0.5\textwidth]{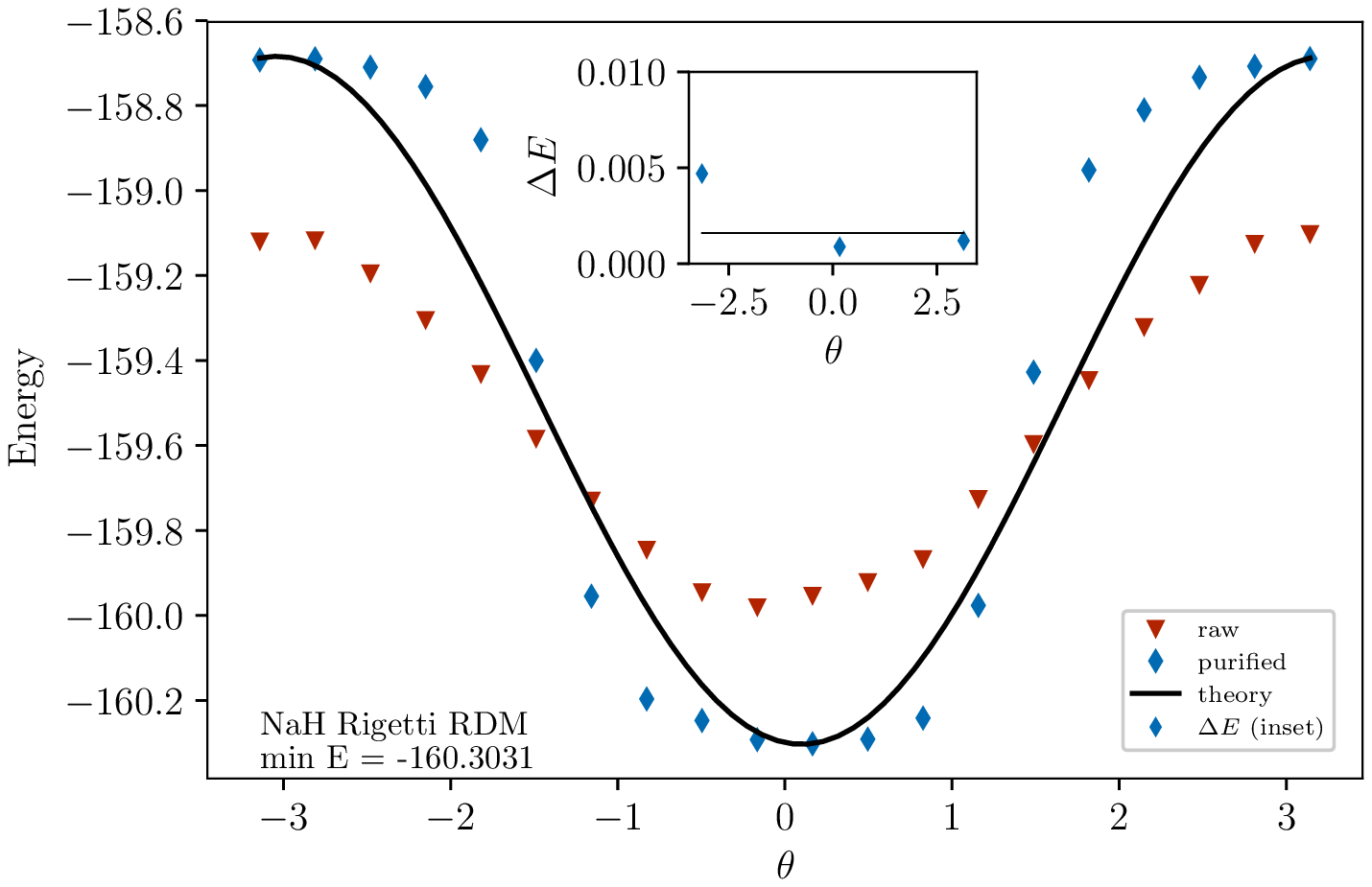} 
\end{center}
\caption{The dependence of the energy as a function of $\theta$ for the ucc-1 ansatz for NaH with and without RDM purification for IBM Tokyo (top) and Rigetti QCS (bottom).}
\label{fig:rdm_rucc_plots}
\end{figure}

First, we considered the reduced, single parameter UCC ansatz (Eq.~\ref{eq:ruccsd}; labeled \emph{ucc-1} in Figure \ref{fig:uccsd}), with readout error correction applied, and at an inter-atomic distance close to the equilibrium configuration of the molecule ($R=1.914388, 2.473066, 2.319238 \si{\angstrom}$ for NaH, RbH, and KH, respectively).
Since this is a 1-parameter problem, we swept this parameter from $[-\pi,\pi]$ and used a cubic spline interpolation to compute the optimal angle. These parameter sweeps (from the IBM QPU) are shown in Figure \ref{fig:plots}.
Each energy point is the average of 8192 shots, and error bars are smaller than the markers (on the order of .001). We then computed the energy at the optimal parameter twice (to accumulate statistics over 16384 shots) with an increasing number of noisy CNOT identity pairs (r=1 is the original circuit, r=3 has each CNOT replaced with 3 CNOTs, etc.). With this data, we performed both linear and quadratic extrapolation to r=0 (the theoretical zero-noise point) to get an estimate of the energy. For this extrapolation, we used the SciPy \texttt{curve\_fit} function to perform a non-linear least squares fit with knowledge of the uncertainty in the data. The linear and quadratic extrapolations are provided by the inset plots for each of the plots in Figure \ref{fig:plots}. In addition, our results are provided in Table \ref{tab:data}. 
\begin{figure}[ht]
\begin{center}
\includegraphics[width=0.5\textwidth]{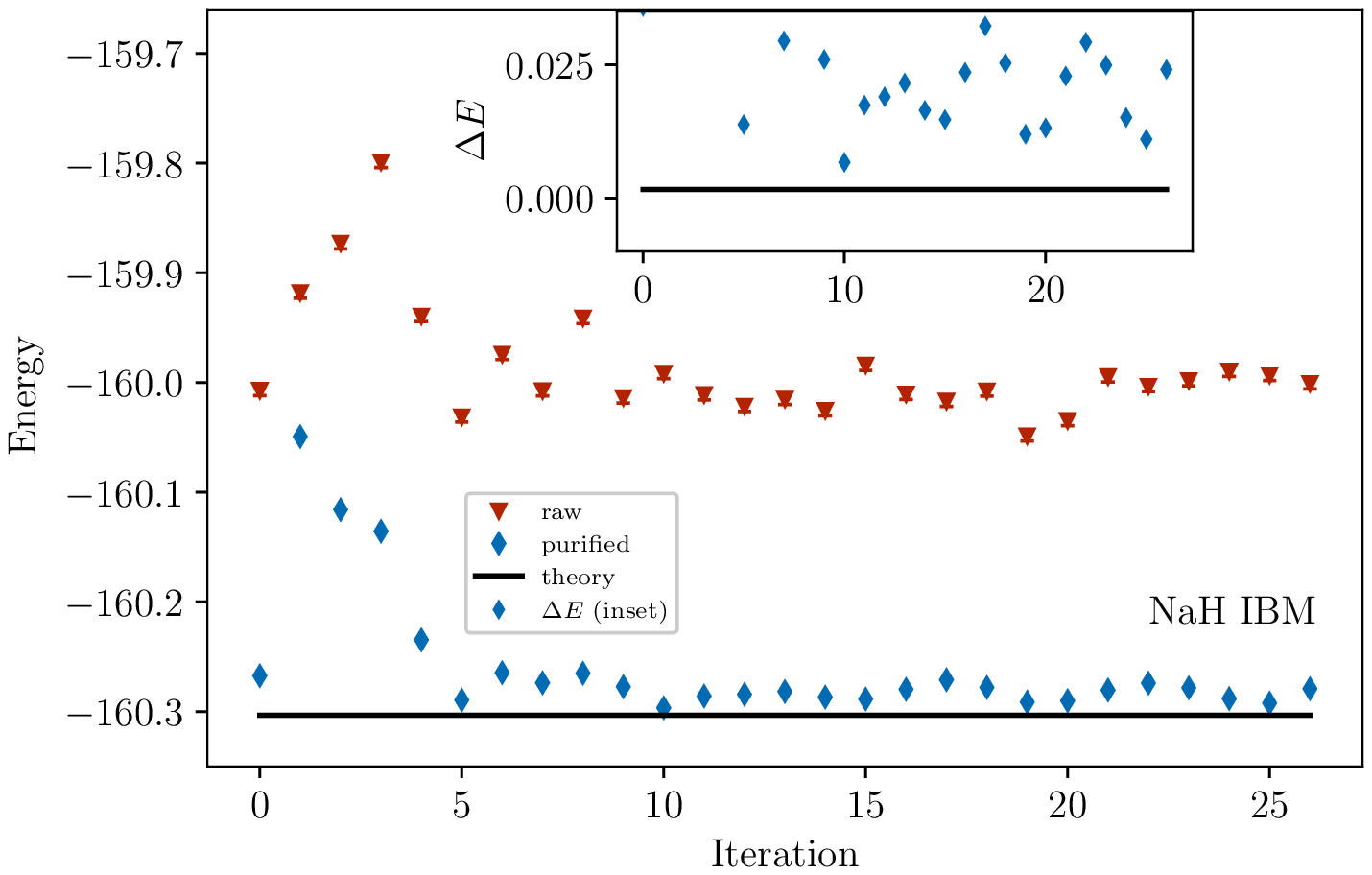} 
\includegraphics[width=0.5\textwidth]{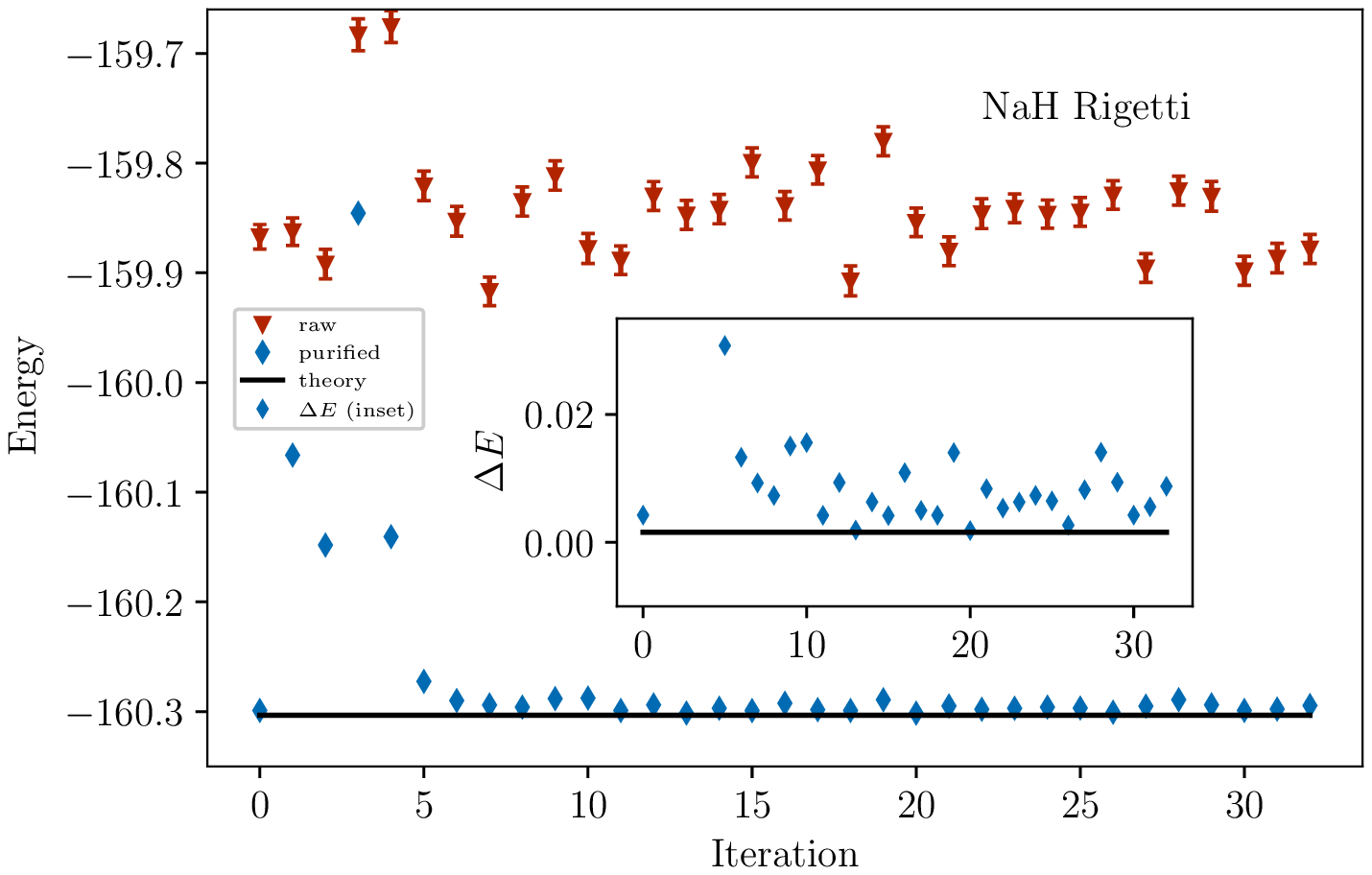} 
\end{center}
\caption{Energy as a function of the optimization iteration for the 3 parameter UCC (ucc-3) ansatz before and after RDM purification on the IBM Tokyo (top) and Rigetti QCS (bottom). The inset plots demonstrate the distance of the computed energy to the FCI energy, with the solid black line denoting chemical accuracy.}
\label{fig:rdm_vqe_jucc_plots}
\end{figure}

We next considered the hardware-efficient ansatz, labeled \emph{hwe} in the table, in which we classically optimized a 20-parameter search space ($d=1$, nearest neighbor entanglement) using the COBYLA method with 30 iterations. We initialized the optimization using a set of angles taken from the simulated optimal set in an effort to avoid local minima. For the \textit{hwe} experiments, we also considered a reduction of the Hamiltonian via application of discrete $Z_2$ symmetries that reduced the qubits needed~\cite{tapering}. The results for these computations are provided in Table \ref{tab:data}.
\begin{figure}[ht!]
\begin{center}
\includegraphics[width=0.5\textwidth]{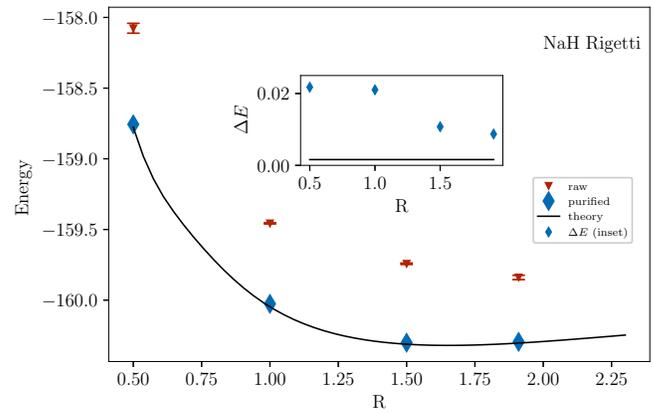} 
\end{center}
\caption{Computed energy as a function of the distance between Na and H for the one parameters UCC (ucc-1) ansatz before and after RDM purification. The inset plot demonstrates the distance of the computed energy from the FCI energy, with the solid black line denoting chemical accuracy. }
\label{fig:rdm_vqe_jucc_nah_surface}
\end{figure}

We found that the \textit{hwe} ansatz was unable to produce results comparable with energies calculated using a classical, full configuration interaction (FCI) method. Symmetry reductions of the total Hamiltonian did improve the energy accuracy, but these results did not get below the theoretical Hartree-Fock energy.
The single parameter UCC ansatz performed better than the hardware-efficient ansatz, despite three additional CNOT gates in the UCC ansatz in an otherwise similar depth circuit. With readout error correction and a quadratic Richardson extrapolation, our benchmark energy metric is comparable to classical FCI results within error bars for all of the reduced 4-qubit Hamiltonians considered here.

As an alternative error mitigation technique, we applied RDM purification while running the benchmarks on both the IBM and Rigetti quantum computers. Figure~\ref{fig:rdm_rucc_plots} shows the results of RDM purification for NaH on Tokyo (top) and Aspen (bottom) using the ucc-1 ansatz. We note that using this error mitigation approach we observe chemical accuracy at the optimal angle on the Rigetti QPU. We also considered the 3-parameter UCC ansatz for NaH and classically optimized via the COBYLA method on both IBM (top) and Rigetti QPUs (bottom) in Figure~\ref{fig:rdm_vqe_jucc_plots}. These computations employed 1000 shots on Rigetti and 8192 shots on IBM. The inset plots demonstrate the difference between purified energy points and the FCI energy, with the solid line denoting chemical accuracy (.0016 Ha). 

Finally, we applied RDM purification to the computation of the NaH energy surface. We chose four inter-atomic distance values, R, and computed $E(R)$ with the one-parameter UCC ansatz. We repeated this five times for each R to gather statistics. Figure \ref{fig:rdm_vqe_jucc_nah_surface} demonstrates the results of this computation, with errors on the order of .01 (smaller than the markers). 

\section{Discussion}
Near-term quantum computation is limited in the total number of operations that can be performed. It is limited in the number of two-qubit entangling gates that can be present in a program before systematic noise reduces the computation to noise. Classical post-processing of the raw experimental data is a requirement to produce usable, classically comparable results. At this nascent stage of quantum computing, a useful computational benchmark that enables cross-platform comparison of various scientific use cases must take these limitations into account.

Here we have proposed and demonstrated such a benchmark - a quantum chemistry benchmark that leverages the robust VQE algorithm, an active space reduction of the electronic structure Hamiltonian, a corresponding reduced and extended unitary coupled cluster ansatz, and various error mitigation strategies. We have demonstrated this primitive benchmark on the IBM Tokyo and Rigetti Aspen machines. Using the computed ground state energy as a metric, we observe results that are comparable with classical simulation results. As new hardware comes online, one can immediately apply this proposed benchmark as a litmus test for the usability of the hardware for quantum chemistry. As existing hardware improves, the active orbitals space  and the ansatz complexity can be increased. We showed that for certain configurations, the benchmark returns chemically accurate results on certain machines. If desired, this metric may be used as a pass/fail criterion to describe a QPU's utility in this very specific task.

We note that RDM purification and the subspace extension to the ansatz were both able to produce improvements over the raw circuit execution. In the case of ansatz extension, this represents a correction to algorithmic error. Whether the computer can obtain the predicted improvement from algorithmic corrections is indirectly a measurement of systematic hardware error. The RDM purification is another indirect measurement of  systematics. The extent to which RDM purification improves our answer tells us about the mixed state of the quantum register.

\section{Conclusion}
We have proposed a novel benchmark of near-term quantum computers. Our approach maps electronic structure calculations from computational chemistry to nascent hardware. We have discussed and demonstrated mechanisms that enable the simulation of larger molecules on noisy, currently available QPUs, as well as various error mitigation strategies that improve computational results. We have proposed a reduced UCC ansatz as the core to a primitive benchmark for near-term hardware, and we have demonstrated this benchmark for alkali metal hydrides like NaH, RbH, and KH. Our work provides a relative baseline for improvements in both hardware and algorithmic approaches for quantum chemistry on quantum computers. 

\section{Acknowledgements}
Z.\ P.\ was supported by an appointment to the Oak Ridge National Laboratory HERE Program, sponsored by the U.S.\ Department of Energy and administered by the Oak Ridge Institute for Science and Education. The authors acknowledge DOE ASCR funding under the Testbed Pathfinder program, FWP number ERKJ332. This research used quantum computing system resources supported by the U.S. Department of Energy, Office of Science, Office of Advanced Scientific Computing Research program office. The authors acknowledge fruitful discussions with E. Dumitrescu. 

\bibliographystyle{apsrev4-1}
\bibliography{references}
\end{document}